\documentclass{aa}
\usepackage{graphics,aanda,amssymb,rotating,psfrag,amsmath,epsfig,array}
\usepackage{color}

\title{ISO observations of the Wolf-Rayet galaxies NGC 5430, NGC 6764, Mrk 309 and VII Zw 19\footnote{Based on observations with ISO, an ESA project with instruments
funded by ESA Member States (especially the PI countries: France,
Germany, the Netherlands and the United Kingdom) with the participation
of ISAS and NASA}}

\titlerunning{ISO observations of several Wolf-Rayet galaxies}

\author{B. O'Halloran\inst{1,2,3} \and B. McBreen \inst{3} \and L. Metcalfe\inst{4}
        \and M. Delaney \inst{3} \and D.Coia\inst{3}}

\offprints{boh@physics.gmu.edu}

\institute{Dunsink Observatory, Castleknock, Dublin 15, Ireland
\and
Dept. of Physics and Astronomy, George Mason University, Fairfax, Virginia, 22030, USA
\and
Physics Department, University College, Belfield, Dublin 4, Ireland
\and
XMM-Newton Science Operations Centre, European Space Agency, Villafranca del Castillo, P.O. Box 50727, 28080 Madrid, Spain.}

\date{Received / Accepted }

\abstract{Observations of four WR  galaxies (NGC 5430, NGC
6764, Mrk 309 and VII  Zw 19) using the Infrared Space Observatory
are presented  here.  ISOCAM maps  of NGC 5430,  Mrk 309 and  NGC
6764 revealed  the location  of star  formation regions in  each
of these galaxies.  ISOPHOT spectral observations  from 4 to 12
$\mu$m detected the ubiquitous PAH bands in the nuclei of the
targets and several of the disk star forming  regions, while LWS
spectroscopy detected [O\,I] and [C\,II] emission  lines from 
two galaxies, NGC 5430 and NGC 6764.

Using a combination of ISO and IRAS flux densities, a dust model
based on the  sum of modified blackbody components  
was successfully fitted to the available data. These models were
then used to calculate new values for the total IR luminosities
for each galaxy, the size of the various dust populations, and the
global SFR.

The  derived flux  ratios, the SFRs, the high  L(PAH)/L(40-120
$\mu$m) and F(PAH 7.7  $\mu$m)/F(7.7 $\mu$m continuum)  values
suggest that  most of these galaxies are home to only  a
compact burst of star formation. The exception is NGC 6764, whose
F(PAH 7.7 $\mu$m)/F(7.7 $\mu$m continuum) value of 1.22 is
consistent  with the presence of  an  AGN,  yet the
L(PAH)/L(40-120 $\mu$m) is more in line with a starburst,  a
finding in line  with a compact low-luminosity AGN dominated
by the starburst.}

\begin{document}

\maketitle   

\section{Introduction}

\subsection{WR galaxies}

Wolf-Rayet (WR) galaxies are  defined as those galaxies whose
integrated spectra contain a broad emission feature at HeII
$\lambda$4686 \cite{conti1991}. This feature has a full-width
half-max (FWHM) of about 10--20\AA~and is a typical signature of
WR stars. Though Seyfert galaxies  and active galactic nuclei
(AGN) often show  a HeII $\lambda$4686 emission line,  WR galaxies
can  be distinguished from  them by their relatively narrow
nebular emission lines.   WR galaxies are found exclusively  among
emission  line (EL) galaxies, where the photoionization of  the
nebular line is stellar in origin, and also possess a  very blue
continuum which is indicative  of a large  population  of  young
hot  massive stars.   The  broad HeII $\lambda$4686  emission
feature is  very prominent  in the spectra of Galactic and LMC WR
stars. By comparing  the luminosity and width of this feature as
it appears in  the spectrum of a WR galaxy, with the corresponding
emission lines from Galactic WR stars, an  estimate of the number
of WR stars  in a WR galaxy  can be made.

The   first  galaxy   in  which   this  WR   feature   was  discovered
\cite{allen1976} was  the blue compact dwarf He  2-10, though NGC~6764
and Mrk~309  were the first objects  to be actually referred  to as WR
galaxies \cite{osterbrock1982}.  The first comprehensive catalogue was
compiled   by   Conti    (1991),   and   included   approximately   40
galaxies.  Since then,  the  number  of known  WR  galaxies has  grown
rapidly to more than 130.  These have been most recently catalogued by
Schaerer  et~al.   (1999).   Many  of  the new  members  of  this
catalogue  show additional features  from WR  stars in  their spectra.
For  example, the broad  emission lines  of NIII  $\lambda$4640 and/or
CIII  $\lambda$4650  as  well  as  CIV  $\lambda$5808  are  among  the
strongest optical lines in WN  and WC stars and are increasingly being
detected.

WR galaxies  are found among  a large variety of  morphological types,
from low-mass  blue compact dwarfs and irregular  galaxies, to massive
spirals and luminous merging IRAS galaxies. There are systems where WR
stars are found in singular giant HII regions (e.g.\ Tol~89 within the
galaxy NGC~5398), in  the nucleus or core (e.g.\  in the barred spiral
LINER NGC~6764), in  knots (e.g.\ at the end of the  bar in the barred
spiral NGC~5430)  and in interacting members of  compact groups (e.g.\
HCG~31A and C) \cite{oha:2002}.  Given  the wide range of morphological
types,   an   age   $\leq10$~Myr    and   high   initial   masses   of
$\mathrm{M}_{ini}$ $\geq35$ $\mathrm{M}_{\odot}$ \cite{maeder1994}, WR
galaxies  are therefore  ideal objects  to study  the early  phases of
starbursts,  determine burst  properties (age,  duration, SFR)  and to
constrain parameters (i.e.\ slope and upper mass cut-off) of the upper
part of the initial mass  function.  Conversely studies of the stellar
populations in super star clusters frequently formed in starbursts and
WR galaxies \cite{conti1994,meurer1995}  can also place constraints on
stellar  evolution models  for massive  stars (e.g\  at  extremely low
metallicities) which are inaccessible in the Local Group.

\subsection{Why observe WR galaxies with ISO?}

Observations of four out of six WR galaxies which were part of the
ISO WRHIIGAL program \cite{oha:2000,oha:2002} are presented in
this paper, along with  supplemental archival observations from
the IRGAL\_1 (PI: Tonaka), GALXISM (PI: Smith), MPEXGAL1 (PI:
Genzel) and WMFP15\_A (PI: Wozniak) programs.  The main goal of
the original WRHIIGAL program was to systematically investigate a
sample  of WR galaxies in order to try to understand what  induces
such a massive burst  of star formation in these   systems,
and  to   investigate  the  mid   and  far-IR characteristics of
such objects. The fact that WR galaxies  are roughly a coeval
sample makes comparisons with other WR  and starburst galaxies
important.  Given the importance  of the mid-infrared region of
the  spectrum for  the physics  of star formation  and for
exploring  the links between  massive star formation in galaxies
and AGN,   acquisition   of   data    in this wavelength range was
crucial. However, no other instruments prior to the launch
of ISO could provide spatially and spectrally resolved
mid-infrared data on a  sample of WR and starburst galaxies -  ISO
data was crucial in  opening up  this important area  of research,
such as those obtained using mid-IR data from surveys of starburst
and blue compact galaxies in the ISO WRHIIGAL and HAROA programs
\cite{steel:1996,oha:2000,oha:2002,met:2005,oha:2005}. The
ISOCAM, ISOPHOT and LWS observations provided maps and spectra of
the galaxies in one, or in combinations of, the Unidentified
Infrared Bands (UIB) and dust or nebular line emission.  They
enabled a determination of the spatial relationship between  the
optical star forming regions and the infrared  emission  from  the
young WR  and starburst regions, allowing a discrimination of the
relative contribution to the infrared flux from  dust, nebular
emission and polycyclic aromatic hydrocarbon (PAH) molecules. The
advent  of the Spitzer Space Telescope, optimized to conduct
research in  a similar wavelength range but  with greater spatial
and  spectral resolution, will further open up  this research
field over the forthcoming years.

\subsection{The sample of WR galaxies}

Four galaxies displaying WR features were observed, as part of a
wider survey of  starburst and  WR galaxies (the WRHIIGAL and
HAROA programs)
\cite{steel:1996,oha:2000,oha:2002,met:2005,oha:2005} by the
Infrared Space Observatory, are discussed in  this section. Table
1 lists the basic properties of each galaxy.

\begin{table}

\caption{Characteristics and  parameters of the WR  galaxy sample. The
four   columns  give  the   galaxy  name,   alternative  designations,
morphological   type  and   distance  in   Mpc,  assuming   H$_{0}$  =
75\,km\,s\,$^{-1}$\,Mpc$^{-1}$.}

\begin{flushleft}
\fontsize{7pt}{9pt}\selectfont
\begin{tabular}[h]{cccc}
\hline\hline \noalign{\smallskip} Galaxy & Other designations & Morphology &
Distance   \\   &   &   &   [Mpc]  \\
\hline
{\bf NGC~5430}  & Mrk~799 & SB(s)b HII & 39.8 \\ \\
{\bf NGC~6764} & UGC~11407 & SB(s)bc; LINER Sy2 &  32.4 \\ \\
{\bf  Mrk~309} & IV Zw 121  & Sa Sy2 HII & 174.1 \\  \\
{\bf VII Zw 19} & PGC~015803 & --  & 65.2 \\
\noalign{\smallskip}\hline
\end{tabular}
\end{flushleft}

\end{table}

\subsubsection{NGC 5430}

NGC 5430  (Mrk 799)  is a nearby  S-shaped barred spiral  galaxy (SBb)
located at  a distance  of 39.8  Mpc.  Keel (1982)  detected the
broad 4650\AA~emission  feature in the knot southeast  of the nucleus,
at the end of the bar.   This knot is an extremely luminous HII region
\cite{keel1982,keel1987}  though   of  unusually  low   luminosity  at
4.885~GHz  \cite{condon1982}, possibly  indicating an  anomalously low
supernova rate.

\subsubsection{NGC 6764}

At a distance of 32.4~Mpc, NGC~6764 is a nearby S-shaped barred
spiral  galaxy  (SBbc),  somewhat  similar  to  NGC~5430.   Originally
considered to be  a Seyfert 2 galaxy by  Rubin et~al.  (1975), Heckman
(1980) later classified NGC~6764 as a LINER on account of its large OI
$\lambda$6300/OIII  $\lambda$5007  ratio  of $\sim0.3$.   This  galaxy
contains a nuclear stellar optical continuum source which extends over
$\sim1.6''$ \cite{rubin1975}.

\subsubsection{Mrk 309}

At     a    distance    of  174.1~Mpc, Mrk~309 is  the most distant object in
the sample. It was identified as  a galaxy with a bright UV continuum,
noticeable  H$\alpha$ emission  and  possible Seyfert  characteristics
\cite{markarian1971}.

Mrk 309 appears in the  sample of 212 emission-line galaxies extracted
from   the    Universidad   Complutense   de    Madrid   (UCM)   lists
\cite{zamorano1994}     and was described as  a  Wolf-Rayet
dominated nucleus galaxy with various HII regions outside the nucleus.
The R-band image  of Mrk 309 presented by  Vitores et~al. (1996) shows
that the southern arm is more  prominent than the northern and ends in
a knot approximately $12''$ from the nucleus.

\subsubsection{VII Zw 19}

VII~Zw~19 appears in  the list of WR galaxies  discovered by Kunth and
Joubert (1985) from a search among a sample of 45 blue EL galaxies.  A
broad emission band between  4600\AA~and 4711\AA~was the criteria used
to decide that  this EL galaxy is also a WR  galaxy. Beck (2000) notes
that the  radio and H$\alpha$ emission  from this galaxy have  the same basic
structure of a very strong central source embedded in a weak envelope.
The central non-thermal  source appears to be unresolved  and is about
800  times  more   luminous  than  Cas  A.   Recent   MERLIN  and  VLA
observations \cite{beck2004} note that VII  Zw 19 resembles M82 in its
radio and infrared spectrum, however the starburst region of VII Zw 19
is twice the size and twice as  luminous as that of M82.  VII Zw 19 is
situated at a distance of 65.2  Mpc.

We present ISOCAM, ISOPHOT and LWS observations of these  galaxies
as part of  an observing  program  consisting of  several galaxies
exhibiting Wolf-Rayet  signatures.   The  observations  and  data
reduction  are presented  in Sect.  2.   The results  are
contained  in Sect.   3 and discussed in Sect.  4.  The
conclusions are summarized in Sect.  5.

\section{Observations and Data Reduction}

The ISO \cite{kes2003} observations were obtained using the mid-infrared
camera  ISOCAM  \cite{blom2003},  the   spectrometric  mode  of  the  ISO
photopolarimeter   ISOPHOT  \cite{laureijs2003a}  and   the  medium-resolution
grating    mode   of    the   long    wavelength    spectrometer   LWS
\cite{gry2003}.   The  astronomical  observing template  (AOT)
used numbers and the observing log  for the ISO observations is presented
in Table 2.

\subsection{ISOCAM}

The  observations had the
following configuration:  1.5, 3  and 6$''$ PFOV  (for Mrk  309
LW2/LW3,  NGC 5430  \& NGC  6764 LW2/LW3  and LW10 observations
respectively), integration time of  2.1~s and $\sim$100 readouts
(excluding discarded stabilisation readouts). The diameter of the
point spread  function (PSF)  central maximum  at the  first Airy
minimum is 0.84 x $\lambda$($\mu$m) arcseconds. The FWHM is about
half that amount. A discussion of the data reduction steps for the CAM data can be
found in O'Halloran et al. (2002).

\subsection{PHT-S}

 PHT-S consists of a dual  grating spectrometer with a resolving power
of 90 \cite{laureijs2003b}.  Band SS covers the range 2.5 - 4.8
$\mu$m, while  band SL  covers the  range 5.8  - 11.6  $\mu$m. Two
different chopping modes were used. The NGC  5430 and Mrk 309
PHT-S spectra were obtained   by  pointing   the  $24''\times24''$
aperture   of  PHT-S alternatively towards the  peak of the
emission (for  512 seconds) and then  towards two  background
positions  off the  galaxy  (256 seconds each), using the ISOPHOT
focal plane chopper.  The NGC 6764 and VII Zw 19  PHT-S spectra
were  obtained on  the other  hand by  operating the PHT-S
aperture in rectangular  chopping mode. The satellite pointed to a
position between  the source  and an  off-source position,  and
the chopper  moved then  alternatively between  these two
positions.  The source   was  always   in  the   positive  beam in
the  spacecraft Y-direction. The calibration  of the spectra was
performed  by using a spectral response  function derived from
several  calibration stars of different brightness observed in
chopper mode \cite{acostapulido2000}. The relative spectrometric
uncertainty  of the PHT-S spectrum is about 20\% when comparing
different parts of the spectrum that are more than a  few microns
apart.   The absolute  photometric uncertainty  is $\leq$
10\% for bright  calibration sources. A discussion of the data reduction steps for the PHOT data can be
found in O'Halloran et al. (2002).

\subsection{LWS}

 The LWS spectra  were obtained with the aperture  centered on the 6.7
$\mu$m ISOCAM map for NGC 5430 and on the nuclear region of the galaxy
for NGC  6764. It  was assumed  that for both  objects the  source was
completely  included in the  beam of  the LWS  instrument, so  that no
extended-source  correction  was  necessary.   The  beam  of  LWS  was
slightly  elliptical and its  FWHM varied  between 65$''$  and 85$''$,
depending on  wavelength and direction  (Swinyard et al.   1998).  The
grating was  fully scanned 6  times over the entire  wavelength range.
A spectral sampling interval of 4 was employed to give 4 spectral
points per resolution element in each of the scans.  All the datasets
were first  processed with version  10 of the LWS  pipeline processing
software,  OLP  V10.1 \cite{gry2003}.   The  LWS Interactive  Analysis
package LIA \cite{iso-sidher-1997esa} was then used to further process
the output of the standard pipeline.  The nominal, fixed, dark current
values, as  determined from  the dedicated measurements  in revolution
650 \cite{iso-swinyard-2000a} were subtracted from the data. The data
was rebinned to one point per LWS detector, employing a scan-averaging
method  described  by  Sidher  et~al.  (2000),  yielding  10  bandpass
continuum  estimates spanning  the LWS  spectral range,  and are
plotted in Fig 3.

\section{Results}

\begin{figure*}
\centering
\includegraphics*[width=13cm]{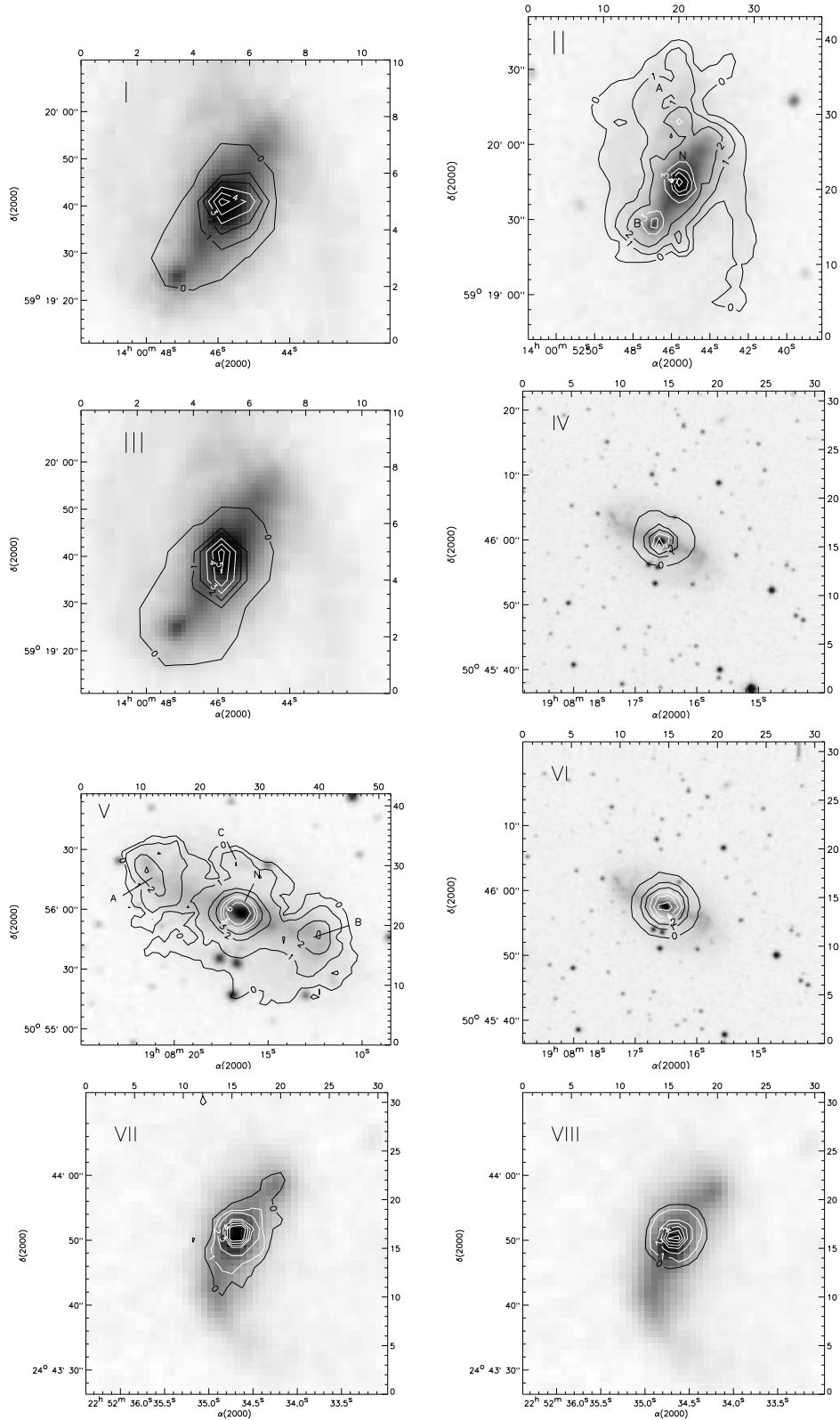}

\caption {ISOCAM maps of NGC  5430 ({\it (I)} at 6.7 $\mu$m, {\it(II)}
at 12 $\mu$m,  {\it(III)} at 14.3 $\mu$m), NGC  6764 ({\it(IV)} at 6.7
$\mu$m, {\it(V)} at  12 $\mu$m, {\it(VI)} at 14.3  $\mu$m) and Mrk 309
({\it(VII)} at  7.7 $\mu$m, {\it(VIII)} at 14.3  $\mu$m), all overlaid
on DSS images.  For  NGC 5430, the contour levels in the  order 0 to 5
are as  follows (in units of  mJy/arcsec\(^{2}\))~:{\it(I)} 0.7, 1.5,
2.3, 3.2, 3.9, 4.8; {\it(II)}  1.9, 3.5, 5.0, 6.5, 8.0, 9.5;
{\it(III)}  0.7, 2.1,  3.9, 6.1,  8.7,  11.4. For  NGC 6764,  the
contour  levels in  the  order 0  to 5  are  as follows  (in units  of
mJy/arcsec\(^{2}\))~: {\it(IV)} 0.6, 1.9, 3.1, 4.4, 5.7, 7.0; {\it(V)}
1.0, 2.9, 4.9, 6.9, 8.8, 10.8; {\it(VI)}  0.6, 2.3, 4.3,  7.5, 11.2,
16.1.  For Mrk  309, the  contour levels in  the order  0 to 5  are as
follows (in  units of mJy/arcsec\(^{2}\))~: {\it(VII)}  0.2, 0.5, 0.9,
1.2, 1.5, 1.9;  {\it(VIII)} 0.2, 0.6, 1.3, 2.0, 2.8,  4.0. The top and
right-hand scales refer to local pixel number scales.}

\end{figure*}

\subsection{ISOCAM maps}
\subsubsection{NGC 5430}

Deconvolved 6.7, 12 and 14.3 $\mu$m maps of NGC 5430 overlaid on a DSS
image are  presented in Figs.  1(I), (II) and (III)  respectively.  It
can clearly be  seen that the ISOCAM contours  follow the general form
of the  galaxy, especially  on the 12  $\mu$m map.  In  addition three
sources were clearly  detected in the higher resolution  12 $\mu$m map
and are labelled A, B and N,  though source B was also detected at 6.7
and  14.3 $\mu$m.   The source  B coincides  with the  highly luminous
southeast HII region  and contains the majority of  the WR population.
The weaker  source A is probably  associated with a  less luminous HII
region  in NGC~5430.  The  nucleus of  NGC~5430 is  denoted N  in Fig.
1(II).  The derived fluxes for NGC~5430 are given in Table 3.

\subsubsection{NGC 6764}

Deconvolved  6.7, 12  and 14.3  $\mu$m  maps  of NGC  6764
overlaid on  a DSS image  are presented in  Figs. 1(IV), (V)  and (VI).
Four infrared sources were detected on the higher resolution 12 $\mu$m
map and are  labelled A, B, C and  N.  The weaker sources A  and B are
probably associated  with star  formation regions in  the arms  of the
galaxy, while  C may also  be a star  formation region or  an external
galaxy. The derived ISOCAM fluxes are given in Table 3.

\subsubsection{Mrk 309}

The 7.7 and 14.3 $\mu$m maps  of Mrk~309 are presented in Figs. 1(VII)
and 1(VIII).  Due to the low signal  to noise ratio of  the 7.7 $\mu$m
image,  deconvolution was  not reliable.  The nucleus  of  Mrk~309 was
detected in  both maps, though more  weakly at 7.7$\mu$m  than at 14.3
$\mu$m. At low flux levels  the 7.7 $\mu$m contours follow the outline
of the  galaxy, while  the 14.3 $\mu$m  emission is more  compact. The
ISOCAM infrared  fluxes are given in  Table 3.

\subsubsection{VII Zw 19}

No observations of VII Zw 19 were performed using ISOCAM.

\subsection{PHT-S spectra}

\subsubsection{NGC 5430}

The PHT-SL spectrum of the  nuclear region of NGC~5430 is presented in
Fig. 2a.   The main  PAH bands  at 6.2, 7.7,  8.6 and  11.3$\mu$m were
easily  detected,  along  with  the [ArII]  ~6.99~$\mu$m  feature.   Two
additional PHT-SL spectra  of the regions labelled A  and B were also
obtained and are presented in Figs. 2b and 2c. In both spectra the PAH
bands plus [ArII] at  6.99~$\mu$m were well detected, along with
a blend of  features at 10.6~$\mu$m  in region A. The
fluxes for the identified features are given in Table 4.

\subsubsection{NGC 6764}

Three PHT-SL spectra of the nucleus and regions A and B were obtained
and are presented in Figs. 2d-2f.  The main PAH bands at 6.2, 7.7, 8.6
and 11.3  $\mu$m were  easily detected in  the nuclear  region, though
only weakly detected at star formation region A. The spectrum of region
B was very noisy and  no features were reliably identified. The fluxes
for the identified features are given in Table 5.

\begin{figure} \resizebox{\columnwidth}{!}
{\includegraphics*{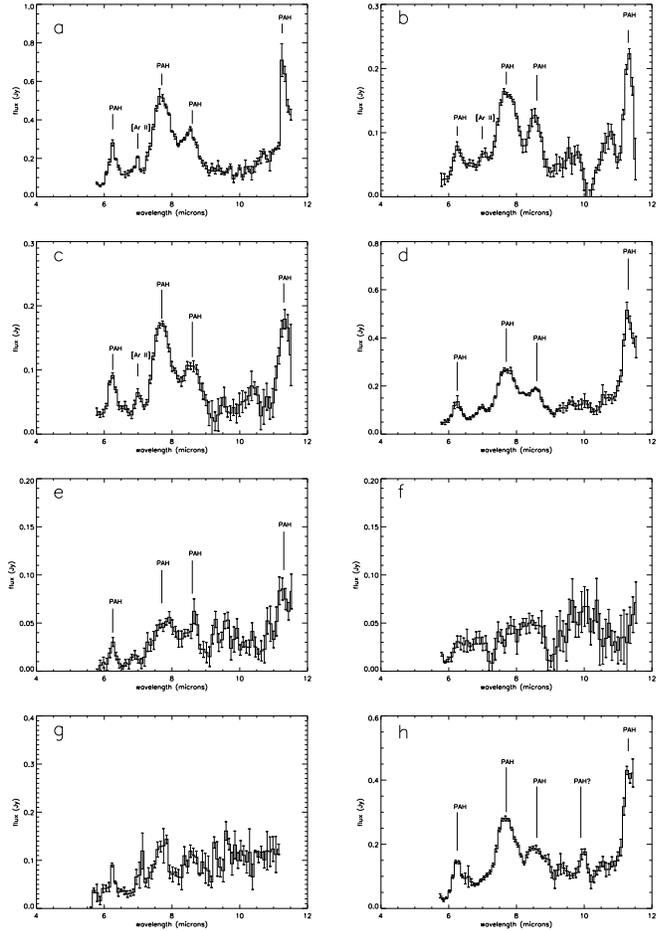}} \caption{PHT-SL spectra
of  NGC 5430 (a: the nucleus, b: region A , c: region B),  NGC 6764  (d: the nucleus, e: region A, f: region B), Mrk  309 (g)  and VII  Zw 19
(h). The PAH features and ionic lines are indicated.}
\end{figure}

\begin{figure} \resizebox{\columnwidth}{!}
{\includegraphics*{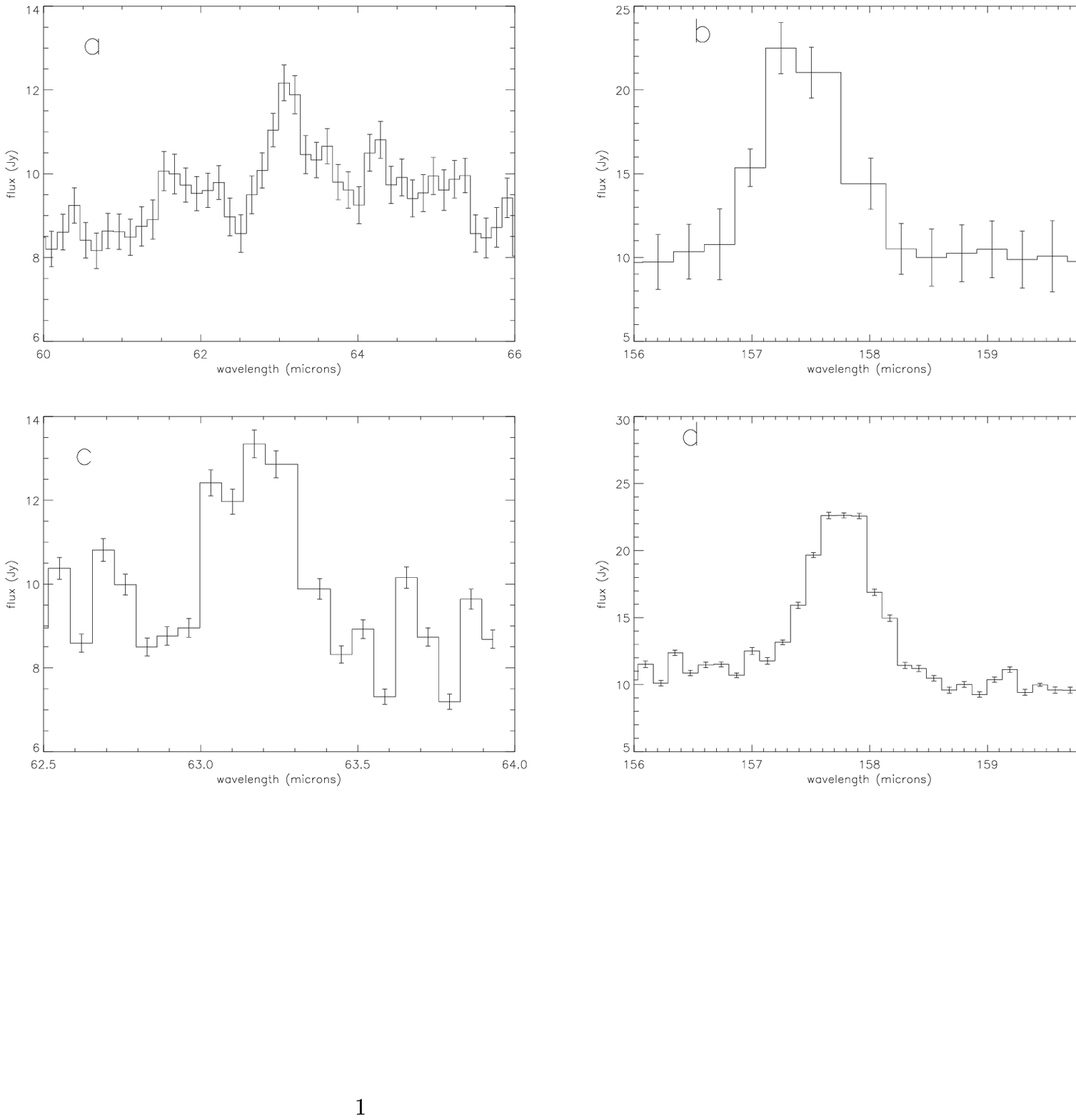}}
\caption{LWS detections of the [OI] 63$\mu$m and 158 $\mu$m lines from NGC 5430 (a,c) \& NGC 6764 (b,d).}
\end{figure}

\subsubsection{Mrk 309}

The PHT-SL spectrum  of Mrk~309, shown in Fig. 2g  was very noisy with
no reliable detections  of the PAH emission bands,  and thus flux
determinations were not made for features in this spectrum.

\subsubsection{VII Zw 19}

The PHT-SL spectrum  of the nucleus of the  VII~Zw~19 is presented in
Fig. 4h.   The main  PAH bands  at 6.2, 7.7,  8.6 and  11.3~$\mu$m are
easily  detected, along  with  [ArII] at ~6.99~$\mu$m and a  possible S(3)  pure  rotational line,
$\upsilon$  =   0-0,  of   molecular  hydrogen  at   9.7~$\mu$m.  The feature at 10.65 $\mu$m may
be a blend of features. The identified line features
and line fluxes are given in Table 5.

\subsection{LWS spectra}

The LWS  spectra for the far-infrared lines [OI] at 63\,$\mu$m and [CII]
at 158\,$\mu$m for both  NGC 5430 and NGC
6764  are presented in  Fig. 3,  while the  line fluxes are  listed in
Table 6. The continuum source strength, or applicable upper-limits,
recorded in the ten LWS detectors by binning data across the
spectral range of each detector are listed in Table 7.

\section{Discussion}

\subsection{Previous surveys}

\subsubsection{NGC 5430}

Keel (1982) estimated that  $\sim5\times10^3$ WN7 stars and $\sim10^4$
WC8 stars reside  in this knot as  well as a still larger  number of O
stars.   With improved spectra  Keel (1987)  later claims  that though
HeII $\lambda$4686 and [NIII] $\lambda$4640 are detected, [CIV] is not
and  therefore the  WR stars  in the  knot might  be regarded  as some
species of WN stars.  Keel  (1987) also identifies 26 other ``normal''
HII regions present in NGC~5430  other than the southeast knot and the
nucleus.

According to Keel (1987) the observed star formation in the SE knot is
both sudden  and transient.  The nebula is  expanding, probably driven
by  stellar  winds,  and  dissipating  on a  timescale  of  $\sim10^7$
years. Star formation elsewhere in the galaxy appears to be proceeding
at  a normal rate  and with  a normal  HII region  luminosity function
\cite{keel1987}. The nebular abundances in the SE knot are rather high
for a  position so far out  in a disk or  bar and are  more like those
seen in nuclei. This suggests that this knot might  be a separate object
resulting from the collision of a dwarf irregular galaxy with the disk
and  bar  of NGC  5430.   Furthermore,  the  H$\alpha$ luminosity  and
emission-line  ratios  are quite  comparable  to  those  seen in  such
systems as Mrk 108 \cite{keel1985}.  Using starburst models by Cervino
and  Mas-Hesse  (1994)  and  Leitherer  and  Heckman  (1995),  Contini
et~al. (1996) estimated the age of  the starburst in the SE knot to be
3~Myr and 4~Myr respectively.  For the nucleus the estimates are 8~Myr
and 9~Myr respectively.  These estimates are certainly compatible with
the scenario of separate star  formation mechanisms in the nucleus and
SE knot of NGC 5430.

\subsubsection{NGC 6764}

Osterbrock  and Cohen  (1982)  detected WR  emission  features in  the
spectrum  of the  nucleus of  this narrow  emission line  galaxy.   From  the  equivalent
widths of  the WR emission features at  4650\AA, Osterbrock and Cohen (1982)  estimate that the
number  of WN  and  WC stars  is  $5.0\times10^4$ and  $9.0\times10^4$
respectively.   Confirmation  of  the  existence of  the  WR  features
detected  by   Osterbrock  and  Cohen   (1982)  was  made   by  Eckart
et~al. (1996), who claim that the number of WO stars is negligible and
WN stars dominate the WR star population.  Detection of the broad [CIII]
$\lambda$5696 and [CIV] $\lambda$5808  lines by Kunth and Contini (1999)
indicate the existence of some WC stars. Eckart et~al. (1996) estimate
that there are  a total of 3600 WR stars  contributing to the ionizing
flux in the  nucleus of NGC 6764. These authors also  note that the WR
feature is spatially extended and most of it originates in the $1.6''$
diameter nuclear  optical continuum source.  Using the  same method as
for  NGC~5430,  Contini  et~al.   (1996)  estimates  the  age  of  the
starburst  in the  nucleus  of NGC~6764  to  be 5.0  Myr  and 6.5  Myr
depending on  which models are  used \cite{cervino1994,leitherer1995}.
The  presence of  WR  stars \cite{osterbrock1982}  indicates that  the
starburst is very young ($\leq6$~Myr) and that many massive stars were
born during the burst. Eckart et~al.  (1996) conclude that the nucleus
of  NGC~6764 has  recently  undergone or  is  currently undergoing  an
intense  starburst with  a  characteristic timescale  of  a few  times
$10^7$ years.

\subsubsection{Mrk 309}

Arkelian  et~al.   (1972)  noted  the presence  of  diffuse  H$\alpha$
emission, as well as  the lines NII $\lambda$$\lambda$6548, 6583.  The
presence  of  these lines  were  later  confirmed  by Afanasev  et~al.
(1980).  Osterbrock and  Cohen (1982) detected a blend  of three broad
emission  features at  4650\AA~in  the nucleus  of  this galaxy:  NIII
$\lambda$4640, CIV  $\lambda$4660 and HeII  $\lambda$4686. These lines
are  slightly narrower than  in NGC  6764 and  hence a  slightly lower
proportion  of  the  continuum  radiation  at  4650\AA~comes  from  WR
stars.  However, Conti  (1991)  argues that  the  emission feature  at
4660\AA~is attributable  to FeIII rather  than CIV, and  is consistent
with the deficiency in the strength of the CIII $\lambda$4650 emission
feature as  noted by Osterbrock  and Cohen (1982).   CIV $\lambda$5808
and CIII $\lambda$5696 emission, attributed to WC stars, has also been
tentatively  detected   \cite{osterbrock1982}.   From  the  equivalent
widths of  the WR emission features at  4650~\AA~these authors deduced
that 9$\%$ of the continuum radiation at this wavelength comes from WR
stars  and hence  calculated  the number  of  WN and  WC  stars to  be
approximately $0.9\times10^4$ and $1.5\times10^4$ respectively.

With the exception of the WR features at 4650\AA, Osterbrock and Cohen
(1982) point out that the emission  line spectra of Mrk 309 is typical
of  low-ionization  galaxies  with  gas  in  their  nuclei  apparently
photoionized by early-type stars, as in HII regions. Indeed Mazzarella
and  Balzano  (1986)  places   Mrk~309  in  the  HII  spectral  class.
Osterbrock and  Cohen (1982) concluded that  the number of  O stars in
Mrk~309 is comparable to the number of WR stars and hence massive star
formation must have occurred recently.

\subsection{Dust components and luminosities}

\begin{figure} \resizebox{\columnwidth}{!}
{\includegraphics*{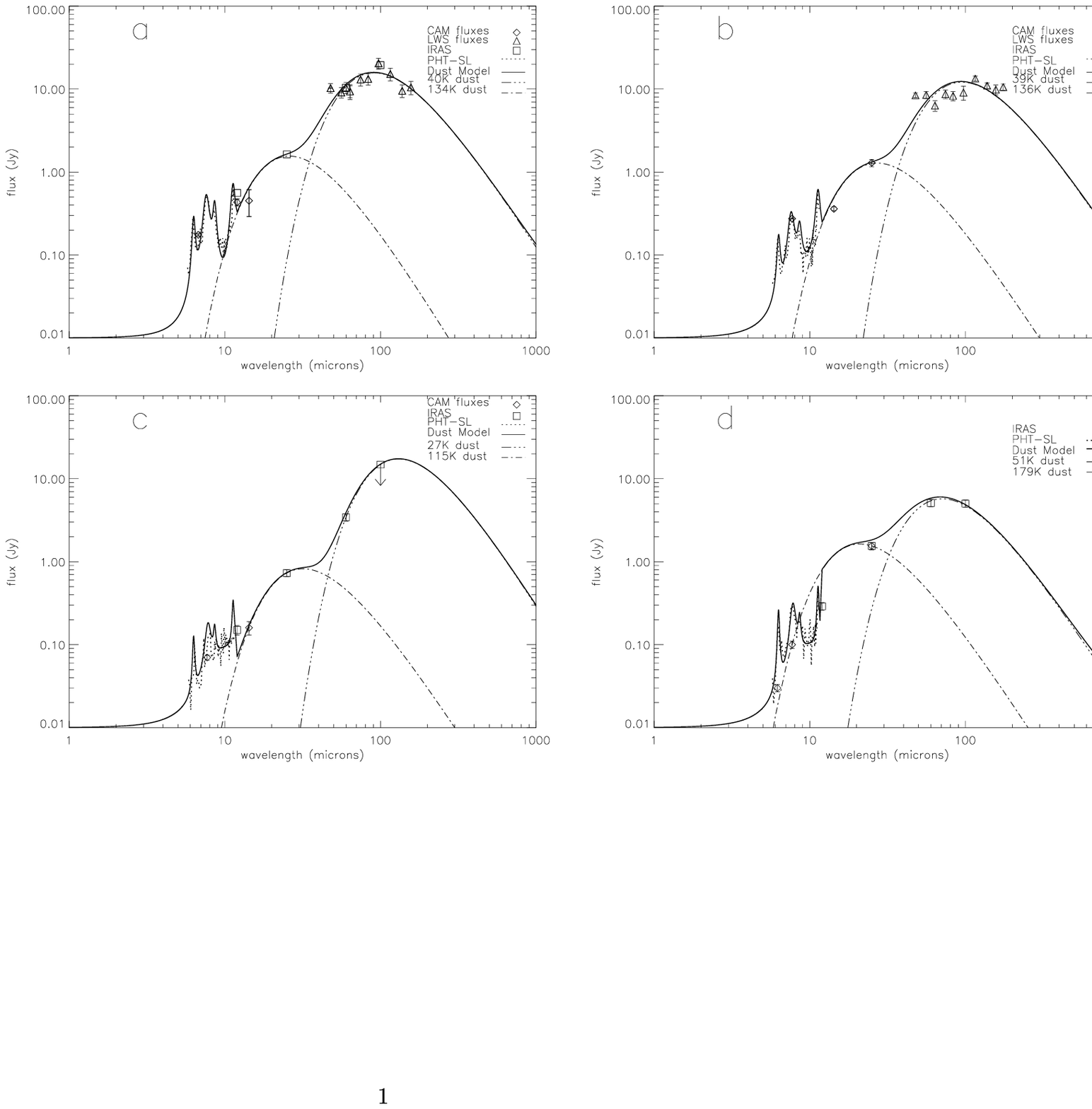}}
\caption{Spectral energy distribution of (a) NGC 5430,  (b) NGC
6764 (c) Mrk 309 and (d) and VII Zw 19 using ISO and  IRAS flux
data, including the key for the different symbols.  The models are
described in the text.}
\end{figure}

There  is growing  evidence for  the existence  of  several
components within   the   dust   distribution   of   galaxies 
\cite{klaas2001}. Specifically  this  can  be  divided
into  warm  dust  components associated  with  star formation
regions and  a  spatially  extended distribution of cold dust.
Warm dust emission is generally associated with very small dust
grains which are  heated by the  single-photon absorption
process    to   temperatures    up   to    several hundred Kelvin
\cite{desert1990},  and  are not  in  thermal equilibrium with
their environment \cite{calzetti2000}.  Emission from very small
dust grains account for the characteristics of the mid-IR at
wavelengths $\leq$ 40 $\mu$m.   The cold dust  emission is
associated with  classical large grains,  emitting at wavelengths
in  excess of  80 $\mu$m  in thermal equilibrium  with their
environment  \cite{calzetti2000}. The  large grains  are heated by
the  normal interstellar  radiation field.  These large dust
grains  account  for  practically all  the emission longward  of
80 $\mu$m in  galaxies.

To model the dust  continuum at mid and far-infrared wavelengths,
modified blackbody functions comprised of a greybody component were
fitted  to  the  IRAS and  ISO  fluxes  for  the  galaxies in  the  WR
sample. The spectral energy  distributions of NGC~5430, NGC 6764, Mrk
309  and VII  Zw 19  using  IRAS, ISOCAM,  PHT-SL and  LWS fluxes  are
presented in Figs.  4a-d.  The dust model, denoted  by the solid curve
was  fitted to  the data  and contains  two separate  dust populations
along with  the PAH bands \cite{boulanger1998}: a  warm dust component
at 135-179 K and a cooler dust component at 39-64 K and each component
is indicated in Figs. 4a-d.

For VII Zw 19 at $\lambda$ $\geq$ 300$\mu$m, it is noticeable that the
emission is  flattening, possibly due to  the presence of  a very cold
dust  component.  However  overall,  the results  are consistent  with
similar modelling by Siebenmorgen et al. (1999).

To determine the mass of  each dust component, several parameters need
to  be determined.  The  luminosity of  each component  was determined
from  the integrated  fluxes of  each  greybody, along  with the  dust
temperature and  then a  dust mass for  each component  was determined
\cite{klaas2001}.   Using this  method, the  IR luminosities  and dust
component masses  in similar starburst  galaxies such as  the Antennae
and NGC  6240 were determined  in order to  check the validity  of the
derived values \cite{klaas1997}.

For each  target, the  luminosity for the  galaxy, the  dust component
luminosity and the  derived dust masses are given in  Table 8. For the
warm and cool dust components, the derived values are quite similar to
starbursts such  as the Antennae, NGC 6240  \cite{klaas1997}, NGC 1741
\cite{oha:2002}  and  Mrk  297   \cite{met:2005}.

\subsection{Current star formation}

\subsubsection{Diagnostics using ISOCAM data}

In order to determine  the current state of star formation within
our sample, diagnostics such as the 14.3/6.75 $\mu$m flux ratio can be
used  as they  act as standard  ratio diagnostics  for this  type of
investigation  \cite{vigroux1999,helou1999}.  Since  the 14.3  $\mu$m
flux is  dominated primarily by  emission from very small  dust grains
and the 6.75 and 7.7 $\mu$m fluxes are dominated by PAHs, the 14.3/7.7
$\mu$m  ratio provides a  diagnostic similar  to the  14.3/6.75 $\mu$m
ratio,  allowing   a  determination  of  the  current   state  of  the
starburst. These diagnostic  ratios generally decrease as interactions
develop and  starbursts age \cite{vigroux1999,helou1999}  - the highly
ionizing  O stars  in the  burst die  off, thus  no  longer destroying
nearby PAHs,  plus emission from  nearby dust heated by  massive stars
also decreases. For the three galaxies with available ISOCAM data, the
derived ratios are given in Table 9. The 6.7 $\mu$m flux for Mrk
309 was synthesized from the  PHT-SL data, a technique used previously
when   faced   with   the    lack   of   6.75   $\mu$m   ISOCAM   data
\cite{oha:2000,oha:2002}.  The obtained  14.3/7.7 14.3/6.75  and
14.3/7.7  $\mu$m  ratios  were  indicative  of  strong  ongoing  star
formation, with ratios generally well  above 2, similar to those found
in earlier works for strong starbursts \cite{vigroux1999}.

\subsubsection{Star formation rate}

Using the calibrations in Kennicutt (1998) and Wilke et al. (2004), we
can calculate the  global star formation rate for  the sample from the
derived  IR luminosities.  Global SFRs  of 9.6,  5.0, 24.1  and 12.6
M$_{\odot}$ yr$^{-1}$ were calculated for  NGC 5430, NGC 6764, Mrk 309
and  VII Zw  19  respectively.  These strong  levels  of ongoing  star
formation   are  consistent   with  the   values  from   the  infrared
diagnostics.

\subsection{Do any galaxies within the sample harbour an AGN?}

\subsubsection{Mid-IR diagnostics}

Historically,  2  galaxies  of   the  sample  have  been
suspected  of harbouring AGN.  Mrk  309 was identified as a galaxy
with a bright UV continuum  and noticeable  H$\alpha$
emission  by Markarian  (1971) who noted the possibility of  a
Seyfert nucleus.  Afanasev (1980) reported H$\alpha$ and [NII]
$\lambda$6548, $\lambda$6583  and stated that Mrk 309 may be  a
Seyfert 2 galaxy.  NGC  6764 has been classified  as a LINER
galaxy  based  on  optical  spectroscopy  \cite{osterbrock1982}.
NIR spectroscopy  of the  nuclear region \cite{schinnerer2000}
reveals that  star  formation is confined to two regions, one with
a radius less than 100 pc containing the WR stars. NGC 6764  also
shows strong variations in ROSAT X-ray flux density  by  at  least
a  factor   of  2  on  timescales  of  7  days
\cite{schinnerer2000}, suggesting the presence  of a compact AGN
with R $\sim$ 10$^{3}$ AU.  No  previous survey has suggested that
NGC 5430 or VII Zw 19 harbours an AGN.

In  order to  test  the  possibility  that the  central  region
harbours an  AGN, diagnostic  tools using ISO  data are  available to
probe the nature of the  activity within the central starburst region.
The ratio of the integrated PAH luminosity and the 40 to 120 $\mu$m IR
luminosity  \cite{lu1999} discriminates  between  starbursts, AGN  and
normal  galaxies. The  lower the  ratio,  the more  likely the  galaxy
harbours an AGN, due to high very small dust grain emission powered by
AGN  longward  of 10  and  shorter  than  50 $\mu$m,  plus  PAH
destruction near the AGN  \cite{vigroux1999}.  Similarly, the ratio of
the 7.7 $\mu$m PAH flux to  the continuum level at this wavelength can
provide  a  measure  of  the  level of  activity  within  the  nucleus
\cite{genzel1998,clavel2000,laureijs2000},  as a high  ratio indicates
the lack  of an AGN excited  dust component. The  derived ratio values
are given in Table 9  for each galaxy.  Given the high L(PAH)/L(40-120
$\mu$m) and  F(PAH 7.7 $\mu$m)/F(7.7 $\mu$m continuum)  values, it can
be stated that most of these galaxies are home to only a compact burst
of star formation  and that an AGN is not  required within the central
bursts. - values of $\leq$  1.5 and 0.06 would be indicative of a
strongly dominant  AGN component.  The  exception is NGC  6764, whose
F(PAH 7.7 $\mu$m)/F(7.7 $\mu$m  continuum) value of 1.22 is consistent
with the presence  of an AGN, yet the  L(PAH)/L(40-120 $\mu$m) is more
in line with  a starburst, a finding which suggests  the presence of a
compact AGN dominated by a strong starburst component.

\subsubsection{Far-IR diagnostics}

Atomic oxygen  and ionized  carbon are the  principal coolants  of the
gaseous interstellar medium via their  fine structure lines in the far
infrared  (FIR).   In particular,  [OI]  at  63\,$\mu$m  and [CII]  at
158\,$\mu$m  dominate  the cooling  in  the photodissociation  regions
associated with massive stars such  as Wolf Rayets, along with [OI] at
146\,$\mu$m  and [OIII] at  88\,$\mu$m \cite{malhotra1997}.   The [OI]
and [CII]  features are  also produced in  the warm atomic  gas behind
dissociative shocks,  in HII  regions or in  photodissociation regions
(PDRs),  while  [OIII] is  more  associated  with denser  environments
within HII regions \cite{malhotra1997,braine1999}.

The [OI] 63\,$\mu$m and [CII]  158\,$\mu$m lines were well detected in
both  NGC  5430  and NGC  6764  and  allowed  a determination  of  the
L$_{CII}$/L$_{FIR}$  and (L$_{CII}$+L$_{OI}$)/L$_{FIR}$  ratios, which
could  be used  to  probe the  nature  of the  environment within  the
galaxy. For NGC 5430, values of L$_{CII}$/L$_{FIR}$ = $1.9 \times
10^{-3}$  and (L$_{CII}$+L$_{OI}$)/L$_{FIR}$  =  $1.1 \times  10^{-2}$
were determined, given an IRAS flux ratio F(60\,$\mu$m)/F(100\,$\mu$m)
= 0.55.   For NGC  6764, values of  L$_{CII}$/L$_{FIR}$ =  $2.9 \times
10^{-3}$  and (L$_{CII}$+L$_{OI}$)/L$_{FIR}$  =  $1.9 \times  10^{-2}$
were determined, given an IRAS flux ratio F(60\,$\mu$m)/F(100\,$\mu$m)
=  0.53. The   L$_{CII}$/L$_{FIR}$  and  (L$_{CII}$+L$_{OI}$)/L$_{FIR}$
ratios are  consistent with
those  from  other  starburst  galaxies,  given  an  IRAS  flux  ratio
F(60\,$\mu$m)/F(100\,$\mu$m)  =  0.66  \cite{malhotra1997,braine1999},
though  for higher dust  temperatures and  star formation  rates these
ratios decrease \cite{malhotra1997}.

\section{Conclusions}

Observations of  four of  a sample of  WR galaxies using  the Infrared
Space Observatory  were presented  in this paper.  ISOCAM maps  of NGC
5430, Mrk  309 and  NGC 6764 revealed  the location of  star formation
regions in each galaxy, while  ISOPHOT spectral observations from 4 to
12  $\mu$m detected  the ubiquitous  PAH bands  in the  nuclei  of the
targets, and several of the disk star forming regions. Strong [OI] and
[CII] lines were detected in LWS spectra of NGC 5430 and NGC 6764.

Using a combination of ISO and IRAS flux densities, a dust model based
on  the sum  of two  modified blackbody  components  were successfully
fitted to  the available data.  The modified  blackbody functions were
comprised   of   a   $1/\lambda$   component  with   temperatures   of
$\sim$40--64~K  and $\sim$140--180~K depending  on the  galaxy.  These
models accounted for  the far-infrared emission and were  then used to
calculate new values for the total IR luminosities for each galaxy and
the size  of the various  dust populations.  For the  dust components,
the  derived  values are  quite  similar  to  starbursts such  as  the
Antennae and  Mrk 297.

The high  ISOCAM flux  ratios are indicative  of strong,  ongoing star
formation  in these  galaxies.  The high  L(PAH)/L(40-120 $\mu$m)  and
F(PAH 7.7 $\mu$m)/F(7.7 $\mu$m  continuum) values also state that most
of these galaxies  are home to only a compact  burst of star formation
and  that an AGN  is  not required  within  the  central bursts.  The
exception however  is NGC 6764,  whose F(PAH 7.7  $\mu$m)/F(7.7 $\mu$m
continuum) value  of 1.22 is consistent  with the presence  of an AGN,
yet the  L(PAH)/L(40-120 $\mu$m) is more  in line with  a starburst, a
finding in line with a compact AGN dominated by the starburst.

\begin{acknowledgements}
The ISOCAM data presented in this paper was analyzed using `CIA',
a joint development by the ESA Astrophysics Division and the
ISOCAM Consortium.  The ISOCAM Consortium is led by the ISOCAM PI,
C.  Cesarsky, Direction des Sciences de la Matiere, C.E.A.,
France.  The ISOPHOT data presented in this paper was reduced
using PIA, which is a joint development by the ESA Astrophysics
Division and the ISOPHOT consortium. The ISO Spectral Analysis
Package (ISAP) is a joint development by the LWS and SWS
Instrument Teams and Data Centers. Contributing institutes are
CESR, IAS, IPAC, MPE, RAL and SRON.

\end{acknowledgements}

\clearpage
\begin{sidewaystable*}
\begin{center}
\caption{Log of the ISO observations  for the four galaxies.  The
ten columns list the object, the TDT number (a unique identifier
of an ISO observation),  the proposal name, the AOT  number
(which identified  the observing  mode used),  the  filter,   the
wavelength  range  ($\Delta\lambda$),  the reference wavelength
of the filter, the FOV, the duration  of the observation, and the center of
the instrument  field of view in right ascension and declination
respectively.} \fontsize{6pt}{10pt}\selectfont
\begin{tabular}[h]{cccccccccccccc}
\hline\hline
Target  &  TDT & Proposal & AOT   &  Filter/Mode   &  $\Delta\lambda$   & $\lambda_\mathrm{ref}$  & FOV  & MxN raster &  $\Delta$ step & BmSw & Duration  & Right Ascension  & Declination  \\
 & &  & & ($\mu$m) & ($\mu$m) & (arcsec) & (arcsec)  &  & dM (arcsec) dN(arcsec) & (arcsec) & (sec) & (RA) &  (DEC) \\
\noalign{\smallskip}
\hline
\noalign{\smallskip}
\\ {\bf  NGC 5430}

 & IRGAL\_1 & 33700518 & CAM01 & LW2   & 5.0--8.0    & 6.75 & 384 x 384  & 3x3 & 96  96 & N/A  & 393  & 14\(^{h}\) 00\(^{m}\) 45.6\(^{s}\)  & 59$^{\circ}$ 19$'$ 44.0$''$ \\
 & IRGAL\_1 & 33700518 & CAM01 & LW3   & 12.0--18.0  & 14.3 & 384 x 384  & 3x3 & 96  96 & N/A  & 300  & 14\(^{h}\) 00\(^{m}\) 45.6\(^{s}\)  & 50$^{\circ}$ 19$'$ 44.0$''$ \\
 & WRHIIGAL & 51400519 & CAM01 & LW10  & 8.0--15.0   & 12.0 & 168 x 144  & 4x3 & 96  96 & N/A  & 1022 & 14\(^{h}\) 00\(^{m}\) 47.4\(^{s}\)  & 59$^{\circ}$ 19$'$ 26.5$''$ \\
 & WRHIIGAL & 51400215 & PHT40 & SS/SL & 2.5--11.6   & --   & 24 x 24    & N/A & N/A & tri150 & 1132 & 14\(^{h}\) 00\(^{m}\) 44.6\(^{s}\)  & 59$^{\circ}$ 19$'$ 57.2$''$ \\
 & WRHIIGAL & 51400313 & PHT40 & SS/SL & 2.5--11.6   & --   & 24 x 24    & N/A & N/A & tri150 & 1132 & 14\(^{h}\) 00\(^{m}\) 45.8\(^{s}\)  & 59$^{\circ}$ 19$'$ 43.6$''$ \\
 & WRHIIGAL & 51400414 & PHT40 & SS/SL & 2.5--11.6   & --   & 24 x 24    & N/A & N/A & tri150 & 1132 & 14\(^{h}\) 00\(^{m}\) 47.4\(^{s}\)  & 59$^{\circ}$ 19$'$ 26.5$''$ \\
 & IRGAL\_1 & 51400417 & LWS01 & --    & 43.0--190.0 & --   & 101        & N/A & N/A &  & 2124 & 14\(^{h}\) 00\(^{m}\) 45.6\(^{s}\)  & 59$^{\circ}$ 19$'$ 43.9$''$ \\
 & IRGAL\_1 & 51400416 & LWS01 & --    & 43.0--190.0 & --   & 101        & N/A & N/A & subtr off & 2126 & 14\(^{h}\) 00\(^{m}\) 45.3\(^{s}\)  & 59$^{\circ}$ 26$'$ 43.9$''$ \\

\\ {\bf NGC 6764}

 & WMFP15\_A & 69301328 & CAM03 & LW2   & 5.0--8.0    & 6.75 &  48 x 48  & N/A & N/A & 600 & 2438 & 19\(^{h}\) 08\(^{m}\) 16.4\(^{s}\) & 50$^{\circ}$ 55$'$ 59.6$''$ \\
 & WMFP15\_A & 69301328 & CAM03 & LW3   & 12.0--18.0  & 14.3 &  48 x 48  & N/A & N/A & 600 & 2438 & 19\(^{h}\) 08\(^{m}\) 16.4\(^{s}\) & 50$^{\circ}$ 55$'$ 59.6$''$ \\
 & WRHIIGAL & 75301710 & CAM01 & LW10  & 8.0--15.0   & 12.0 &  168 x 132 & 4x4 & 24  12 & N/A & 2126 & 19\(^{h}\) 08\(^{m}\) 16.4\(^{s}\) & 50$^{\circ}$ 55$'$ 58.5$''$  \\
 & MPEXGAL1 & 63801718 & PHT40 & SS/SL & 2.5--11.6   & --   &  24 x 24   & N/A & N/A & rect060 & 1132 & 19\(^{h}\) 08\(^{m}\) 16.4\(^{s}\) & 50$^{\circ}$ 55$'$ 59.4$''$  \\
 & WRHIIGAL & 74501017 & PHT40 & SS/SL & 2.5--11.6   & --   &   24 x 24  & N/A & N/A & rect060 & 1132 & 19\(^{h}\) 08\(^{m}\) 12.8\(^{s}\) & 50$^{\circ}$ 55$'$ 48.8$''$ \\
 & WRHIIGAL & 11601781 & PHT40 & SS/SL & 2.5--11.6   & --   &   24 x 24  & N/A & N/A & 180 & 1132 & 19\(^{h}\) 08\(^{m}\) 21.5\(^{s}\) & 50$^{\circ}$ 55$'$ 56.8$''$ \\
 & GALXISM  & 30200964 & LWS01 & --    & 43.0--190.0 & --   &  84        & N/A & N/A &  & 1920 & 19\(^{h}\) 08\(^{m}\) 16.3\(^{s}\) & 50$^{\circ}$ 55$'$ 59.5$''$ \\

\\ {\bf Mrk 309}

 & WRHIIGAL & 55100531 & CAM03 & LW3   & 12.0--18.0  & 14.3 &   48 x 48  & N/A & N/A & 60 & 1010 & 22\(^{h}\) 52\(^{m}\) 34.7\(^{s}\) & 24$^{\circ}$ 43$'$ 49.5$''$ \\
 & WRHIIGAL & 55100433 & CAM03 & LW6   & 7.0--8.5    & 7.7  &   48 x 48  & N/A & N/A & 60 & 1220 & 22\(^{h}\) 52\(^{m}\) 34.7\(^{s}\) & 24$^{\circ}$ 43$'$ 49.5$''$ \\
 & WRHIIGAL & 55100306 & PHT40 & SS/SL & 2.5--11.6   &  --  &   24 x 24  & N/A & N/A & tri090 & 1132 & 22\(^{h}\) 52\(^{m}\) 34.7\(^{s}\) & 24$^{\circ}$ 43$'$ 49.0$''$. \\

\\ {\bf VII Zw  19}
 & WRHIIGAL & 63702104 & PHT40 & SS/SL & 2.5--11.6   & --   &   24 x 24  & n/a & n/a &  rect030 & 1132 & 04\(^{h}\) 40\(^{m}\) 39.3\(^{s}\) & 67$^{\circ}$ 44$'$ 20.0$''$ \\
\noalign{\smallskip}\hline
\end{tabular}
\end{center}
\end{sidewaystable*}
\clearpage

\begin{table*}

\caption{Fluxes for  the galaxies  observed by ISOCAM. A photometric
uncertainty of 15 \% was assumed.}

\fontsize{10pt}{12pt}\selectfont
\begin{tabular}{ccccc}
\hline\hline

 Reference wavelength $\lambda_\mathrm{ref}$ & 6.7 $\mu$m & 7.7 $\mu$m & 12.0 $\mu$m  & 14.3 $\mu$m \\
 Wavelength range $\Delta\lambda$ & 5.0--8.5 $\mu$m & 7.0--8.5 $\mu$m & 8.0--15.0$\mu$m &  12.0--18.0 $\mu$m \\
 & { Flux [mJy] } & { Flux [mJy] } & { Flux [mJy] } & { Flux [mJy] } \\ \hline

\bf NGC 5430 & 218 $\pm26$ &  -- & 459 $\pm65$ & 441 $\pm68$ \\

\bf NGC~6764 & 160 $\pm24$ &  -- & 310  $\pm47$ & 417 $\pm63$ \\

\bf Mrk~309 &  -- & 74 $\pm11$ & & 160 $\pm24$ \\

\hline
\end{tabular}

\end{table*}

\begin{table*}
\caption{PHT-SL line  intensities for NGC~5430. The  five columns give
the  target, the  line identification,  the reference  wavelength, the
wavelength    range   and    the   integrated    flux   respectively.}
\fontsize{10pt}{12pt}\selectfont
\begin{tabular}{llllll}
\hline\hline Target & Line ID & $\lambda$  & $\Delta\lambda$ & Flux \\ & & {
[$\mu$m]  } &  { [$\mu$m]  } &  ($\times$\,10$^{-15}$\,W/m\(^{2}\)) \\
\hline

{\bf NGC~5430 Nucleus} & PAH 6.2  & 6.2 & 5.8--6.6 & 6.98 $\pm0.45$ \\
& ArII 6.99  & 6.99 & 6.8--7.1 &  2.17 $\pm0.10$ \\ & PAH 7.7  & 7.7 &
7.2--8.2  & 13.84  $\pm0.49$ \\  & PAH  8.6 &  8.6 &  8.2--9.0  & 5.88
$\pm0.43$ \\ &  PAH 11.3 & 11.3 & 11.0--11.6 &  8.88 $\pm0.76$ \\ \\

{\bf NGC 5430 Region  A} & PAH 6.2 &  6.2 & 6.0--6.4 & 1.24  $\pm0.22$ \\ &
ArII 6.99  & 6.99 &  6.8--7.1 & 0.59  $\pm0.14$ \\ &  PAH 7.7 &  7.7 &
7.2--8.2 & 4.14 $\pm0.64$ \\ & PAH 8.6  & 8.6 & 8.4 - 9.1 & 1.09 $\pm$
0.45 \\ &  PAH 11.3 & 11.3  & 11.0--11.6 & 1.61 $\pm0.45$  \\ \\

{\bf NGC 5430 Region B} &  PAH 6.2 & 6.2 & 6.0--6.4 &  1.01 $\pm0.23$ \\ & ArII
6.99 & 6.99 & 6.8--7.1 & 0.51  $\pm0.10$ \\ & PAH 7.7 & 7.7 & 7.2--8.2
& 4.33 $\pm0.45$ \\ & PAH 8.6 & 8.6 & 8.4 - 9.1 & 1.99 $\pm$ 0.39 \\ &
SIV 10.51 & 10.51 & 10.4--10.8 & 0.32 $\pm0.08$ \\ & PAH 11.3 & 11.3 &
11.0--11.6 & 1.42 $\pm0.55$ \\ \hline
\end{tabular}

\end{table*}

\begin{table*}
\caption{PHT-SL line intensities for NGC 6764 and VII Zw 19 determined
assuming a local continuum. The five columns give the target, the line
identification, the reference wavelength, the wavelength range and the
integrated flux respectively.}  \fontsize{10pt}{12pt}\selectfont
\begin{tabular}{llllll}
\hline\hline Target & Line ID & $\lambda$  & $\Delta\lambda$ & Flux \\ & & {
[$\mu$m]  } &  { [$\mu$m]  } &  ($\times$\,10$^{-15}$\,W/m\(^{2}\)) \\
\hline
{\bf NGC~6764  Nucleus} &  PAH  6.2 &  6.2 &  5.8--6.6 &  6.31
$\pm0.40$ \\ & PAH 7.7 & 7.7  & 7.2--8.2 & 8.14 $\pm0.51$ \\ & PAH 8.6
& 8.6 & 8.2--9.0 & 2.82 $\pm0.43$  \\ & PAH 11.3 & 11.3 & 11.0--11.6 &
0.0 $\pm0.79$ \\ \\

{\bf NGC 6764 Region  A} & PAH 6.2 & 6.2 & 6.0--6.4 &
0.87 $\pm0.24$ \\ & PAH 7.7 & 7.7 & 7.2--8.2 & 1.73 $\pm0.44$ \\ & PAH
8.6  & 8.6 &  8.4 -  9.1 &  0.91 $\pm$  0.34 \\  & PAH  11.3 &  11.3 &
11.0--11.6  & 0.67  $\pm0.22$ \\  \\

{\bf VII~Zw~19}  & PAH  6.2 &  6.2 &
5.8--6.6 &  3.12 $\pm0.28$  \\ & ArII  6.99 &  6.99 & 6.8--7.1  & 1.23
$\pm0.12$ \\ & PAH 7.7 & 7.7  & 7.2--8.2 & 8.02 $\pm0.53$ \\ & PAH 8.6
& 8.6 &  8.3--8.9 & 2.55 $\pm0.34$ \\  & PAH? 9.7 & 9.7  & 9.4--10.0 &
2.73 $\pm0.32$ \\  & PAH 11.3 & 11.3 & 11.0--11.6  & 3.55 $\pm0.52$ \\
\hline
\end{tabular}

\end{table*}
\clearpage

\begin{table*}
\caption[]{LWS  line-fluxes  for NGC 5430 and NGC 6764. The  four columns
give  the  species, the  wavelength  of the  line  and  the line fluxes for each galaxy
respectively.}
\begin{flushleft}
\fontsize{10pt}{12pt}\selectfont
\begin{tabular}{lccc}
\hline\hline\noalign{\smallskip}
 Species     &  $\lambda_\mathrm{ref}$ &    NGC 5430 Line Flux  &     NGC 6764 Line Flux \\
             &  ($\mu$m)  &     ($\times$\,10$^{-15}$\,W/m\(^{2}\))             &   ($\times$\,10$^{-15}$\,W/m\(^{2}\))  \\
\noalign{\smallskip}
\hline
\noalign{\smallskip}
$[$O I$]$    &     63.1   &     7.5\,$\pm$\,0.9   &       5.8\,$\pm$\,0.7      \\
$[$C II$]$   &    157.7   &     11.2\,$\pm$\,1.3   &       10.3\,$\pm$\,1.1       \\
\noalign{\smallskip}
\hline
\end{tabular}
\end{flushleft}
\end{table*}

\begin{table*}
\caption[]{Fluxes measured with LWS, binned to yield a single
number per LWS detector.}
\begin{flushleft}
\fontsize{10pt}{12pt}\selectfont
\begin{tabular}{cccc}\hline\hline
\noalign{\smallskip}
Detector label &   $\lambda_\mathrm{ref}$ &  NGC 5430 Flux Density  &  NGC 6764 Flux Density\\
               &  ($\mu$m)                &      (Jy)    &      (Jy)  \\
\noalign{\smallskip}
\hline
\noalign{\smallskip}
     SW1        &  46.8                   &      10.3    &      8.4   \\
     SW2        &  56.0                   &      9.1    &      8.5   \\
     SW3        &  63.7                   &      9.4     &      8.7    \\
     SW4        &  74.4                   &      13.1     &      8.3    \\
     SW5        &  83.3                   &      13.3     &      9.1    \\
     LW1        &  97.2                   &      20.5     &      13.4   \\
     LW2        & 115.7                   &      15.2     &      11.0   \\
     LW3        & 137.8                   &      9.5     &      9.9    \\
     LW4        & 156.9                   &      10.5     &      10.6    \\
     LW5        & 175.2                   &      10.0UL     &    7.7    \\
\noalign{\smallskip}\hline
\end{tabular}
\end{flushleft}
\label{lws-log}
\end{table*}

\begin{table*}
\caption{Dust luminosities and masses}
\begin{flushleft}
\fontsize{10pt}{12pt}\selectfont
\begin{tabular}{ccccc}
\hline\hline

Galaxy & T$_{warm dust}$ & T$_{cool dust}$ & L$_{IR}$ (ISO) & L$_{IR}$
(IRAS)   \\

&  (K)  &  (K)  &  [L$_{\odot}$] &  [L$_{\odot}$]  \\
\noalign{\smallskip}
 \hline\noalign{\smallskip}

{\bf Mrk 309} &  143 & 64 & 1.40 x 10\(^{11}\)  & 3.58 x 10\(^{11}\) \\

{\bf NGC 5430} & 134 & 40  & 5.55 x 10\(^{10}\) & 8.78 x 10\(^{10}\)  \\

{\bf NGC 6764}  & 136 & 39 &  2.85 x 10\(^{10}\) & 3.53  x 10\(^{11}\) \\

{\bf VII Zw 19} & 179 & 51 & 7.35 x 10\(^{10}\) & 1.07 x 10\(^{11}\) \\

\noalign{\smallskip} \hline
\end{tabular}
\end{flushleft}

\end{table*}

\begin{table*}
\caption{Star formation/AGN  diagnostic ratios using CAM  and PHT-S data.}
\fontsize{10pt}{12pt}\selectfont
\begin{tabular}{ccccc}
\hline\hline \\
Galaxy & 14.3/6.75  $\mu$m & 14.3/7.7 $\mu$m & F(PAH 7.7  $\mu$m)/ & L(PAH)/ \\
  & & & F(7.7  $\mu$m continuum) & L(40  --  120$\mu$m)  \\  \hline
{\bf  Mrk  309}  &  2.16 & 2.2 & 5.78 & -- \\
{\bf NGC 5430} & 2.57 & 2.1 & 4.79 & 0.091 \\
{\bf NGC 6764}  & 2.60 & 2.0 & 1.22 &  0.088 \\
{\bf VII Zw 19} & -- & -- & 4.82 & 0.122 \\ \hline
\end{tabular}

\end{table*}


\begin{thebibliography}{}

\bibitem[\protect\astroncite{Abergel et~al.}{1996}]{abe96}
Abergel, A., Bernard, J.P., Boulanger, F. et al., 1996, A\&A , 315, L329

\bibitem[\protect\astroncite{Acosta-Pulido et~al.}{2000}]{acostapulido2000}
Acosta-Pulido, J.A., Gabriel, C., \& Castaeda, H., 2000, Experimental
Astronomy 10, 333

\bibitem[\protect\astroncite{{Afanasev et~al.}}{1980}]{afanasev1980}
Afanasev, V.L., Lipovetskii, V.A., Markarian, B.E. et~al., 1980,
Astrofizika 16, 193

\bibitem[\protect\astroncite{{Allen et~al.}}{1976}]{allen1976}
Allen, D.A., Wright, A.E., \&  Goss W.M., 1976, MNRAS, 177, 91

\bibitem[\protect\astroncite{{Arkelian et~al.}}{1972}]{arkelian1972}
Arkelian, M.A., Dibai, E.A.,\&  Episov, V.F., 1972, Astrofizika 8, 177

\bibitem[\protect\astroncite{Aussel et~al.}{1999}]{aus99}
Aussel, H., Cesarsky, C.J., Elbaz, D.  et~al.,  1999, A\&A, 342, 313

\bibitem[\protect\astroncite{{Beck}}{2000}]{beck2000}
Beck, S.C., 2000, AJ 120, 244

\bibitem[\protect\astroncite{{Beck et~al.}}{2004}]{beck2004}
Beck, S.C., Garrington, S.T., Turner, J.L., \& Van Dyk, S.D., 2004, astro-ph 0406560

\bibitem[\protect\astroncite{Blommaert et~al.}{2003}]{blom2003} Blommaert, J., Siebenmorgen, R., Coulais, A.  et~al.,  2001,   \newblock {ISO Handbook,
Volume  II: The ISO Camera },   ESA, SP-1262,
http://www.iso.vilspa.esa.es/manuals


\bibitem[\protect\astroncite{Boulanger et~al.}{1998}]{boulanger1998}
Boulanger, F., Boissel, P., Cesasrsky, D.  et~al., 1998, A\&A 339, 194

\bibitem[\protect\astroncite{{Braine \& Hughes}}{1999}]{braine1999}
Braine, J. \& Hughes, D.H., 1999, A\&A ~344, 779

\bibitem[\protect\astroncite{Calzetti et~al.}{2000}]{calzetti2000}
Calzetti, D., Armus, L., Bohlin, R.C.  et~al., 2001, AJ 533, 682

\bibitem[\protect\astroncite{{Cervino \& Mas-Hesse}}{1994}]{cervino1994}
Cervino, M., Mas-Hesse, J.M., 1994, A\&A 284, 749

\bibitem[\protect\astroncite{Clavel et~al.}{2000}]{clavel2000}
Clavel, J., Schulz, B., Altieri, B. et~al., 2000, A\&A 357, 839

\bibitem[\protect\astroncite{{Condon et~al.}}{1982}]{condon1982}
Condon, J.J., Condon, M.A., Gisler, G. et~al., 1982, ApJ 252, 102

\bibitem[\protect\astroncite{{Conti}}{1991}]{conti1991}
Conti, P.S., 1991, ApJ 377, 115

\bibitem[\protect\astroncite{{Conti \& Vacca}}{1994}]{conti1994}
Conti, P.S, Vacca, W.D., 1994 ApJL, 423, 97

\bibitem[\protect\astroncite{{Contini}}{1996}]{contini1996}
Contini, T.,
\newblock In: J.M. de~Vreux et~al., eds, {WR Stars in the Framework of Stellar Evolution},
33rd Li\`ege Int. Astroph. Coll., page 619, Li\`ege, 1996.
Universit\'e de Li\`ege.

\bibitem[\protect\astroncite{Delaney et~al.}{2002}]{delaney2002}
Delaney, M. \& Ott, S.\  2002,
http://www.iso.vilspa.esa.es/users/expl\_lib/CAM\_top.html   ``ISOCAM
Interactive Analysis User's Manual'', Version 5.0, SAI/96-5226/Dc.

\bibitem[\protect\astroncite{{Desert} et~al.}{1990}]{desert1990}
Desert, F., Boulanger, F., \& Puget, J.L., 1990, A\&A 237, 215

\bibitem[\protect\astroncite{{Eckart et~al.}}{1996}]{eckart1996}
Eckart, A., Cameron, M., Boller, T. et~al., 1996, ApJ 472, 588


\bibitem[\protect\astroncite{Gabriel}{2002}]{gabriel2002}
Gabriel, C.,  2002, \newblock PHT Interactive  Analysis User
Manual, ESA, http://www.iso.vilspa.esa.es/manuals/

\bibitem[\protect\astroncite{Genzel et~al.}{1998}]{genzel1998}
Genzel, R., Lutz, D., Sturm, E.  et~al., 1998, \newblock ApJ 498, 579

\bibitem[\protect\astroncite{Gry et~al}{2003}]{gry2003}
Gry, C., Swinyard, B.M., Harwood, A. et~al., 2003,  \newblock {ISO Handbook,
Volume  III:   LWS   -   The   Long-Wavelength   Spectrometer},   ESA, SP-1262,
http://www.iso.vilspa.esa.es/manuals



\bibitem[\protect\astroncite{{Heckman}}{1980}]{heckman1980}
Heckman, T.M., 1980, A\&A ~87, 152

\bibitem[\protect\astroncite{Helou}{1999}]{helou1999}
Helou, G., 1999,  In: Cox, P.,  Kessler, M.F. (eds)  {The Universe as seen by
ISO}, ESA SP-427 Vol.2, 797


\bibitem[\protect\astroncite{{Keel}}{1982}]{keel1982}
Keel W.C., 1982, PASJ ~94, 765

\bibitem[\protect\astroncite{{Keel et~al.}}{1985}]{keel1985}
Keel, W.C., Kennicutt, R.C., Hummel, E. et~al., 1985, AJ 90, 708

\bibitem[\protect\astroncite{{Keel}}{1987}]{keel1987}
Keel, W.C., 1987, A\&A ~172, 43

\bibitem[\protect\astroncite{Kennicutt}{1998}]{kennicutt1998}
Kennicutt, R.C., 1998, ApJ 498, 541

\bibitem[\protect\astroncite{Kessler et~al.}{2003}]{kes2003} Kessler, M.F., Muller, T.G., Leech, K.. et~al., 2003,  \newblock {ISO Handbook,
Volume  I:  ISO - Mission \& Satellite},   ESA, SP-1262,
http://www.iso.vilspa.esa.es/manuals


\bibitem[\protect\astroncite{Klaas  et~al}{1997}]{klaas1997}
Klaas, U., Haas, M., Heinrichsen, I. \& Schulz B., 1997, A\&A 325, 21

\bibitem[\protect\astroncite{Klaas et~al.}{2001}]{klaas2001}
Klaas, U., Haas, M., Muller, S.A.H. et~al., 2001, A\&A 379, 823


\bibitem[\protect\astroncite[{Kunth \& Joubert}{1985}]{kunth1985}
Kunth, D., Joubert, M., 1985, A\&A 142, 411

\bibitem[\protect\astroncite{Kunth \& Contini}{1999}]{kunth1999}
Kunth, D., Contini, T., 1999, \newblock In: {Wolf-Rayet stars as
tracing the AGN--starburst connection},  p.~725


\bibitem[\protect\astroncite{Laureijs et~al}{2000}]{laureijs2000}
Laureijs, R.,  Watson,  D., Metcalfe, L., et~al,. 2000, A\&A 359, 900

\bibitem[\protect\astroncite{Laureijs et~al}{2003a}]{laureijs2003a}
Laureijs, R.J., Klaas, U., Richards, P.J.  et~al., 2003, \newblock {ISO Handbook,
Volume  IV:   PHT - The Imaging Photo-Polarimeter},   ESA, SP-1262,
http://www.iso.vilspa.esa.es/manuals


\bibitem[\protect\astroncite{Laureijs et~al}{2003b}]{laureijs2003b}
Laureijs, R.J., Klaas, U., Richards, P.J.  et~al., 2003, \newblock
{ISOPHOT Data Users Manual}, \newblock ESA,
\newblock http://www.iso.vilspa.esa.es/manuals/



\bibitem[\protect\astroncite{{Leitherer \& Heckman}}{1995}]{leitherer1995}
Leitherer, C., Heckman, T.M., 1995, ApJSS 96, 9

\bibitem[\protect\astroncite{{Lemke et~al.}}{1996}]{lemke1996}
Lemke, D.,  Klaas, U., Abolins, J.  et~al., 1996, A\&A 315, 64

\bibitem [\protect\astroncite{Li et~al.}{1998}]{li1998}  Li, W., Modjaz, M., Treffers, R. R., \& Filippenko, A. V.,
1998, IAUC 6850


\bibitem[\protect\astroncite{Lu et~al.}{1999}]{lu1999}
Lu, N.Y., Helou,  G., Silbermann  N. et~al., 1999,  \newblock In:  Cox P.,
Kessler M.F. (eds), {The Universe  as seen by ISO}, ESA SP-427 Vol.2, 929

\bibitem[\protect\astroncite{Malhotra et~al.}{1997}]{malhotra1997}
Malhotra, S., Helou, G., Stacey, G. et~al., 1997, ApJ 491, 27

\bibitem[\protect\astroncite{{Maeder \& Conti}}{1994}]{maeder1994}
Maeder, A., Conti, P.S., 1994, ARA\&A, 32, 227

\bibitem[\protect\astroncite{{Markarian \& Lipovetskii}}{1971}]{markarian1971}
Markarian, B.E, Lipovetskii, V.A., 1971, Astrofizika 7, 511

\bibitem[\protect\astroncite{{Mazzarella \& Balzano}}{1986}]{mazzarella1986}
Mazzarella, J.M., Balzano, V.A., 1986, ApJSS 62, 751


\bibitem[\protect\astroncite{Metcalfe et~al.}{2005}]{met:2005}
Metcalfe, L., O'Halloran, B., McBreen, B. et~al., 2005, submitted to A\&A


\bibitem[\protect\astroncite{Meurer et~al.}{1995}]{meurer1995}
Meurer, G.R., Heckman, T.M., Leitherer, C., et~al., 1995, AJ 110,
2665


\bibitem[\protect\astroncite{O'Halloran et~al}{2000}]{oha:2000}
O'Halloran, B., Metcalfe, L., Delaney, M. et~al., 2000, A\&A 360, 871

\bibitem[\protect\astroncite{O'Halloran et~al}{2002}]{oha:2002}
O'Halloran, B., Metcalfe, L., McBreen B. et~al., 2002, ApJ 575, 747

\bibitem[\protect\astroncite{O'Halloran et~al}{2005}]{oha:2005}
O'Halloran, B., Steel, S., Metcalfe, L. et~al., 2005, submitted to
A\&A

\bibitem[\protect\astroncite{Okumura}{1998}]{okumura1998}
Okumura, K., 1998, ``ISOCAM PSF Report'',
http://www.iso.vilspa.esa.es/

\bibitem[\protect\astroncite{{Osterbrock and Cohen}}{1982}]{osterbrock1982}
Osterbrock D.E., Cohen R.D., 1982, ApJ 261, 64

\bibitem[\protect\astroncite{Ott et~al.}{1997}]{ott1997}
Ott, S., Abergel, A., Altieri, B.  et~al., 1997, \newblock In: Hunt
G., Payne H.  (eds.), { Astronomical Data Analysis Software and
Systems VI}, Vol.  125 of {ASP Conf.  Series}, p.~34


\bibitem[\protect\astroncite{{Rubin et~al.}}{1975}]{rubin1975}
Rubin, V.C., Thonnard, N.C., \& Ford, W.K., 1975, ApJ 199, 31

\bibitem[\protect\astroncite{Siebenmorgen et~al.}{1999}] {isiebenmorgen1999}
Siebenmorgen, R., Krugel, E., Chini, R., 1999, A\&A 351, 495


\bibitem[\protect\astroncite{{Schaerer et~al.}}{1999}]{schaerer1999}
Schaerer, D., Contini, T., \& Pindao, M., 1999, A\&ASS, 136, 35

\bibitem[\protect\astroncite{{Schinnerer et~al.}}{2000}]{schinnerer2000}
Schinnerer, E., Eckart, A., \& Boller, T., 1999, AJ, 545, 205

\bibitem[\protect\astroncite{Sidher et~al}{1997}]{iso-sidher-1997esa}
Sidher, S. D., Swinyard, B. M., Harwood, A. S. et~al., 1997,
\newblock In:  Heras, A.M., Leech, K., Trams, N.R., Perry, M. (eds),
{Proceedings of the first ISO workshop on Analytical Spectroscopy,
Madrid, Spain, 6-8 October 1997}, ESA SP-419, 297

\bibitem[\protect\astroncite{Sidher et~al}{2000}]{iso-sidher-2000}
Sidher, S. D., Griffin, M. J., Davis, G. R. et~al., 2000, Icarus 147, issue 1, 35

\bibitem[\protect\astroncite{{Starck et~al.}}{1998}]{starck1998}
{Starck} J.L., {Murtagh} F., {Bijaoui} A., 1998, in: Image
processing and data analysis. The multiscale approach, Publisher:
Cambridge, UK: Cambridge University Press, 1998, ISBN: 0521590841

\bibitem[\protect\astroncite{Steel et~al.}{1996}]{steel:1996}
Steel S., Smith N., Metcalfe L., Rabbette M., McBreen B., 1996,
A\&A, 311, 721S

\bibitem[\protect\astroncite{{Stevens \& Strickland}}{1998}]{stevens1998}
Stevens, I.R., Strickland, D.K., 1998, MNRAS, 294, 523

\bibitem[\protect\astroncite{Swinyard et~al}{1998}]{iso-swinyard-1998spie}
Swinyard, B.M., Burgdorf, M.J., Clegg, P.E. et~al., 1998,
\newblock In: Fowler, A.M. (ed), {Infrared Astronomical Instrumentation}, Proc. SPIE 3354, 888


\bibitem[\protect\astroncite{Swinyard et~al}{2000}]{iso-swinyard-2000a}
Swinyard, B.M., Clegg, P.E., Leeks, S. et~al., 2000, Experimental Astronomy 10, 157

\bibitem[\protect\astroncite{{Thuan \& Gunn}}{1976}]{thuan1976}
Thuan, T.X., Gunn, J.E., 1976, PASP 88, 543

\bibitem[\protect\astroncite{Vigroux et~al}{1999}]{vigroux1999}
Vigroux, L., Charmandaris, V., Gallais, P.  et~al., 1999, \newblock In: Cox P.,
Kessler M.F.  (eds), { The Universe as seen by ISO}, ESA SP-427 Vol.2, 805

\bibitem[\protect\astroncite{{Vitores et~al.}}{1996}]{vitores1996}
Vitores, A.G., Zamorano, J., Rego, M. et~al., 1996, A\&AS 118, 7

\bibitem[\protect\astroncite{Wilke et~al.}{2004}]{wilke2004}
Wilke, K., Klaas, U., Lemke, D. et al., 2004, A\&A 414, 69

\bibitem[\protect\astroncite{{Zamorano et~al.}}{1994}]{zamorano1994}
Zamorano, J., Rego, M., Gallego, J. et~al., 1994, ApJSS 95, 387


\end{thebibliography}
\end{document}